\documentclass[preprint]{aastex}
\usepackage{emulateapj5}

\shorttitle{Cooling flows or bubbles?}
\shortauthors{McCarthy et al.} 
\journalinfo{astro-ph/0303270}
\submitted{To appear in ApJ Letters (received 01/16/03,
accepted 03/17/03)}

\begin{document} 

\title{On the Relationship between Cooling Flows and Bubbles}

\author{Ian G. McCarthy$^{1}$, Arif Babul$^{1,2}$, Neal Katz$^3$, and Michael L. 
Balogh$^4$}

\affil{$^1$Department of Physics \& Astronomy, University of Victoria, Victoria, BC, 
V8P 1A1, Canada, mccarthy@uvastro.phys.uvic.ca; babul@uvic.ca}

\affil{$^3$Department of Astronomy, University of Massachusetts, Amherst, MA, 01003, USA, 
nsk@kaka.astro.umass.edu}

\affil{$^4$Department of Physics, University of Durham, South Road, Durham, DH1 3LE,
UK, m.l.balogh@durham.ac.uk}

\footnotetext[2]{CITA Senior Fellow}

\begin{abstract} 

A common feature of the X-ray bubbles observed in {\it Chandra} images of some ``cooling 
flow'' clusters is that they appear to be surrounded by bright, cool shells.  Temperature 
maps of a few nearby luminous clusters reveal that the shells consist of the coolest gas 
in the clusters --- much cooler than the surrounding medium.  Using simple models, we 
study 
the effects of this cool emission on the inferred cooling flow properties of clusters.
We find that the introduction of bubbles into model clusters that {\it do not} have 
cooling flows results in temperature and surface brightness profiles that resemble those 
seen in nearby ``cooling flow'' clusters.  They also approximately reproduce the recent 
{\it XMM-Newton} and {\it Chandra} observations of a high minimum temperature of 
$\sim$1-3 keV.  Hence, bubbles, if present, must be taken into account when inferring the 
physical properties of the ICM.  In the case of some clusters, bubbles may account 
entirely for these observed features, calling into question their designation as clusters 
with cooling flows.  However, since not all nearby ``cooling flow'' clusters show 
bubble-like features, we suggest that there may be a diverse range of physical phenomena 
that give rise to the same observed features.

\end{abstract}


\section{INTRODUCTION}

Observations obtained with the {\it Chandra} and {\it XMM-Newton} X-ray Observatories have 
yielded a number of important results that have changed our view of galaxy groups and 
clusters, especially those systems that have been termed ``cooling flow'' clusters$^5$.  
For example, {\it Chandra}'s exquisite spatial resolution has allowed for much more 
detailed analyses of the X-ray surface brightness depressions (referred to as ``bubbles'' 
or ``holes'') discovered in earlier {\it ROSAT} images of several nearby ``cooling flow'' 
clusters (Fabian et al. 2000; Schmidt et al. 2002; Heinz et al. 2002; Blanton et al. 2001; 
2003).  High quality {\it Chandra} data is also responsible for the discovery of many new 
bubbles (or bubble-like features) in a number of other groups and clusters (e.g., McNamara 
et al. 2000; 2001; Schindler et al. 2001; Mazzotta et al. 2002; Johnstone et al. 2002; 
Young et al. 2002; Sanders \& Fabian 2002; Smith et al. 2002).  It now seems that such 
bubbles are a fairly common constituent of ``cooling flow'' clusters.

\footnotetext[5]{The designation {\it ``cooling flow'' cluster} refers to a system that 
has a sharply rising surface brightness profile and a declining temperature profile 
towards the center.  These observational characteristics have typically been interpreted 
as manifestations of an ICM that is 
radiatively cooling on short timescales.  The cooling gas flows inward toward the cluster 
center (hence, the name cooling flow).  When we use the phrase ``cooling flow'' (in 
quotation marks) we are referring to the observational characteristics and not a physical 
model.}

Another important result, derived with {\it XMM-Newton} data, is the lack of spectral 
evidence for gas cooling to temperatures below a few keV (e.g., Peterson et al. 
2001; 2003; Kaastra et al. 2001; Tamura et al. 2001).  Possible explanations for this 
unexpected behavior include heating of the cooling flows by AGN outflows and/or thermal 
conduction, rapid mixing of the low temperature gas, and inhomogeneous metallicity 
distributions in the ICM (e.g., Peterson et al. 2001; Ciotti \& Ostriker 2001; Narayan \& 
Medvedev 2001; Fabian et al. 2002a; 2002b; Churazov et al. 2002; Ruszkowski \& 
Begelman 2002; Kaiser \& Binney 2003; Morris \& Fabian 2003).

The near simultaneous discovery of the connection between bubbles and ``cooling flow'' 
clusters, and the high minimum temperatures in clusters raises the question: are these 
phenomena related?  As we already mentioned, it has been hypothesized that heating by a 
central AGN could quench the cooling flows.  Recent numerical simulations show that 
heating the ICM near the cluster core can also give rise to bubble-like features that 
resemble those seen in the {\it Chandra} images (e.g., Churazov et al. 2001; Quilis et al. 
2001; Brighenti \& Mathews 2002a).  However, it still is not clear {\it how} the AGNs or 
the bubbles they produce could heat up cooling flows, e.g. through shocks, cosmic rays, 
or Compton heating, or whether this heating would be sufficient to offset the radiative 
losses and establish the observed high minimum temperature (see, e.g., Fabian et al. 
2002a; Brighenti \& Mathews 2002b).  We speculate that there could be an even simpler 
connection between the bubbles, ``cooling flow'' clusters, and the high minimum 
temperatures of clusters.
 
A common feature of the X-ray bubbles present in the {\it Chandra} images is that they 
appear to be partially or fully surrounded by cool, bright shells.  In fact, high 
resolution cluster temperature maps of Perseus and A2052 (see Fig. 6. of Schmidt et 
al. 2002; Fig. 10. of Blanton et al. 2003), two nearby X-ray bright clusters which have 
probably yielded the best constraints on bubble properties, reveal that the shells consist 
of the coolest gas in the clusters; much cooler than surrounding ambient medium.  
What are the effects of these bright, cool shells on the inferred cooling flow 
properties of clusters?  It is clear that if the emission from the bubbles is relatively 
important, it will have an impact on both the azimuthally-averaged surface brightness and 
emission-weighted temperature profiles.  Since the cooling flow properties of clusters 
(e.g., the cooling time, mass deposition rate, age and size of the cooling flow) are 
deduced from these profiles, they will also be affected.  To date, however, the effects 
that bubbles have on the inferred properties of gas in the cores of clusters have not 
been studied theoretically or observationally.

In this Letter, we explore how the presence of bubbles affects the surface brightness 
and temperature ($kT_{ew}$) profiles of clusters.  We show that the introduction of 
bubbles into non-cooling flow model clusters results in profiles that closely 
resemble those observed in nearby ``cooling flow'' clusters that clearly contain bubbles 
(but which have not been excised from the analysis of those clusters).  This implies that 
the bubbles have a significant impact on the inferred cooling flow properties of these 
clusters and, in the case of some clusters, may account for the entire ``cooling flow''.

\section{Model Clusters with Bubbles}

To ascertain the effects of bubbles on the general appearance of clusters, we make use of 
analytic ``preheated'' cluster models developed in Babul et al. (2002).  Since an in-depth 
discussion of the models can be found in that study, we give only a very brief description 
here.

The distribution of the dark matter in the model clusters is assumed to be the same as 
that found in recent high resolution numerical simulations.  The intracluster gas, 
preheated to a uniform `entropy' ($\equiv kT_e n_e^{-2/3}$) of 300 keV cm$^2$,  is 
assumed to be in hydrostatic equilibrium within the cluster potential well.  The preheated 
models (with entropy floors $\gtrsim 300$ keV cm$^2$) have been shown to provide an 
excellent match to the observed {\it global} X-ray and thermal Sunyaev-Zeldovich effect 
properties of groups and clusters (Balogh et al. 1999; Babul et al. 2002; McCarthy et al. 
2002; 2003).  A welcome by-product of the high level of preheating is that the cooling 
timescale of the ICM is greater than the age 
of the Universe (for $H_0$ = 75 km s$^{-1}$ Mpc$^{-1}$, $\Omega_m = 0.3$, and 
$\Omega_{\Lambda} = 0.7$ at $z = 0$, which we assume throughout) for groups up to moderate 
mass clusters.  Thus, the complicated effects of radiative cooling and cooling flows 
(which are neglected by the Babul et al. 2002 models) are unimportant for these model 
clusters.  Because there are no cooling flows, it is straightforward to quantify the 
effects of the cool bubble shells on the surface brightness and emission-weighted 
temperature profiles.

For the bubbles, we use the {\it Chandra} images of Perseus and A2052 as a guide.  Each 
model bubble consists of a spherical `cavity' surrounded by a spherical shell$^6$.  Schmidt 
et al. (2002) and Blanton et al. (2001; 2003) argue that any gas filling the `cavities' 
must be hot ($kT_e \gtrsim 20$ keV) and have a low density.  We assume a constant cavity 
temperature of 20 keV.  The density distribution of the cavities is set by requiring that 
they are in pressure equilibrium with the bubble shells and the ambient ICM.  
{\epsscale{1.0}
\plotone{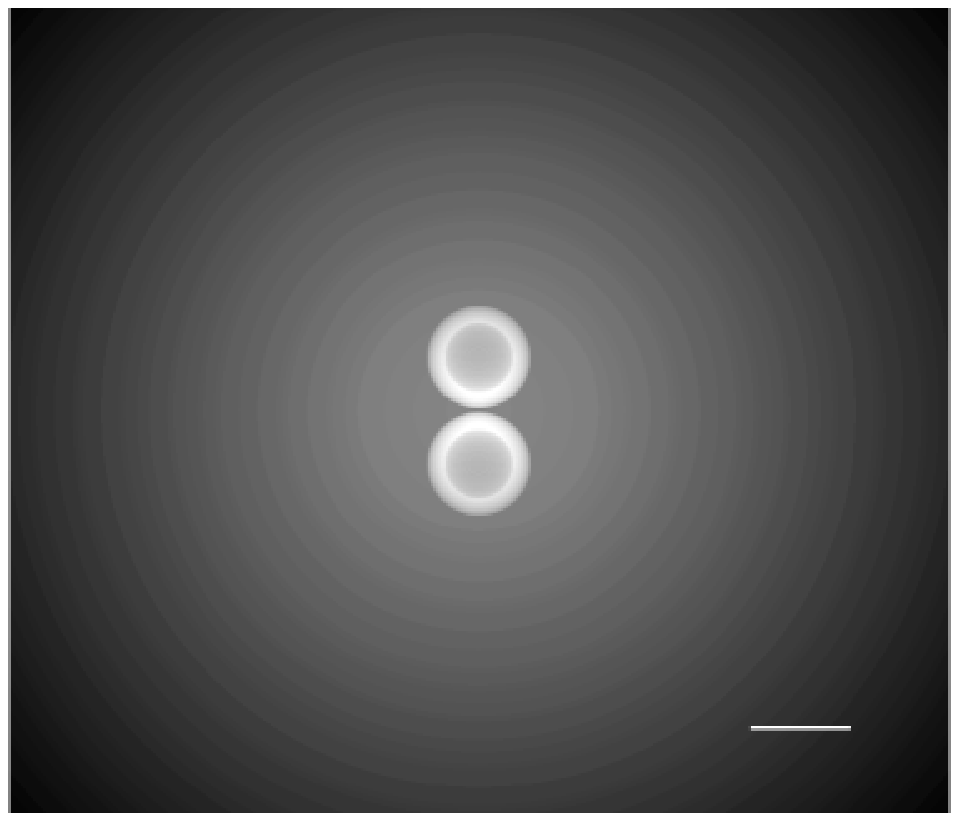}
{Fig. 1. \footnotesize
Bolometric surface brightness map of a typical model cluster.  The surface brightness is
displayed in logarithmic scale.  The solid white line indicates a length of 20 kpc.}}
\vskip0.1in
\noindent
This, combined with the high temperature, insures that the density is quite low and, 
consequently, the cavities are 
X-ray-deficient (as observed).  As expected, use of higher cavity temperatures (i.e., 
lower densities) gives very similar results.  The radius of the cavities is assumed 
to be 7 kpc, approximately the mean value of the bubbles observed in Perseus and A2052 
(scaled to our assumed cosmology).  For the shells, Blanton et al. (2001; 2003) find a 
deprojected temperature of about 1 keV.  We assume this temperature, although changing 
the temperature by up to 50\% does not significantly modify the results (see Fig. 2).  
Again, the density distribution is set by requiring that the shells are in pressure 
equilibrium with the surroundings.  A shell thickness of 3.5 kpc is assumed.  Two of 
these (identical) bubbles are placed near the center of each model cluster.  The bubbles 
are placed in opposite hemispheres with equal distances from the cluster center, and 
perpendicular to the line-of-sight.  We have also experimented with other orientations 
(e.g., bubbles overlapping) but the qualitative results remain generally unaffected.

\footnotetext[6]{For simplicity, we assume that the shells completely surround the cavities, 
even though this does not appear to be the case for all of the observed bubbles.}

A surface brightness map of a typical model cluster with bubbles is displayed in Figure 1.  
As observed, the shells have been significantly `limb-brightened' (especially near the 
cluster center).  With an emission-weighted temperature of $\sim 3$ keV at a projected 
radius of about 50 kpc, beyond the outer radius of the bubble shells, this 
particular model cluster roughly resembles A2052.  

\section{Results}

In Figure 2, we plot the predicted emission-weighted temperature profiles of two model 
clusters.  As expected, the addition of the bubbles with cool shells leads to a decrease in 
the emission-weighted temperature towards the center of the cluster.  The magnitude and 
scale over which the drop occurs, however, is surprising.  The temperature, $kT_{ew}$, 
declines from $\approx 3$ keV to $\approx 2$ keV in the case of the lower mass cluster and 
from $\approx 6$ keV to $\approx 2.5$ keV for the 
{\epsscale{1.0}
\plotone{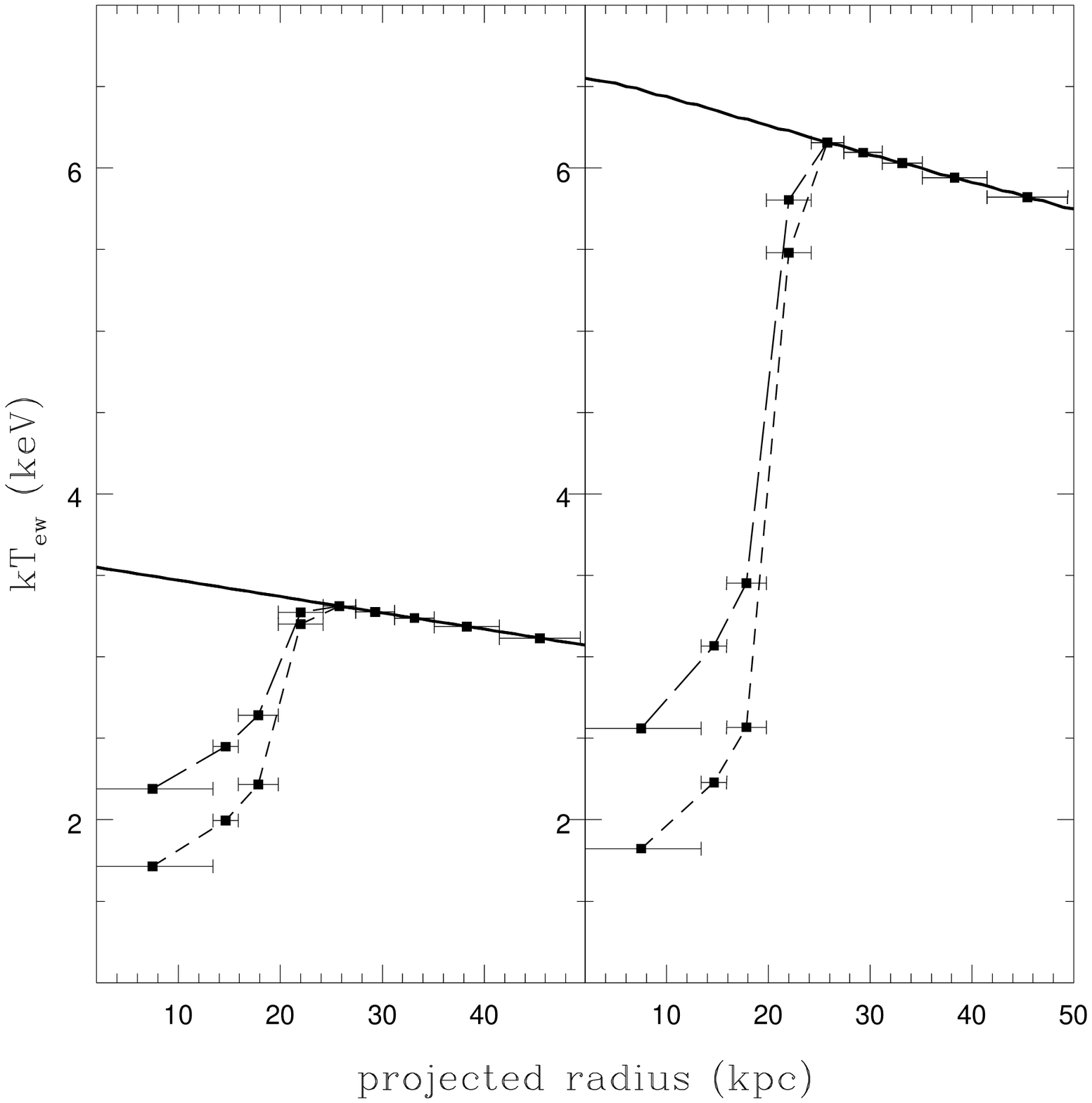}
{Fig. 2. \footnotesize
Predicted emission-weighted temperature profiles.  {\it Left:} Profile of the
cluster displayed in Fig. 1.  {\it Right:}  Profile of a more massive cluster.  The thick
solid lines are the profiles {\it prior} to placing the bubbles in the cluster.  The
long-dashed and short-dashed lines are the profiles assuming shell temperatures of 1 keV
and 0.8 keV, respectively.  The solid squares indicate the radial bins from which the
temperatures were extracted, while the error bars indicate the bin widths (which are
similar to those used in the analyses of Perseus and A2052).}}
\vskip0.1in
\noindent
more massive cluster.  Furthermore, both model clusters 
(in fact, all of the model clusters that we examined) show a slow decline, 
almost a core, in the temperature profile near the very centers of the clusters and the 
minimum temperatures are quite similar ($\sim 2$ keV).  These predicted trends roughly 
match those seen in nearby ``cooling flow'' clusters (that contain bubbles).  This is 
surprising since it implies that the cool shells alone could be entirely responsible for 
the observed temperature dips and the surface brightness peaks (i.e., cooling flows may 
not be necessary for these clusters).  It should be kept in mind that the bubble shells 
have very low masses ($\sim 10^9 M_{\odot}$) and only occupy $\approx 18\%$ 
(combined) of the total volume within the central 21 kpc.  Any mass deposition rates 
inferred from such clusters that do not excise the cool shell emission will grossly 
overestimate the true cooling rate.    

The temperature dips seen in Fig. 2 are obviously confined within the (projected) outer 
radius of the bubble shells (in this case about 21 kpc).  An interesting question, 
therefore, is do the observed temperature gradients in clusters with bubbles extend 
beyond the outer radius of the observed bubbles?  If so, this would immediately imply 
that the bubble shells cannot be {\it solely} responsible for the gradients.  A close 
examination of Fig. 2 of Blanton et al. (2001) suggests that the gradient of A2052 does, 
indeed, begin very near the outer edge of the bubble shells.  Similar, although 
somewhat less clear-cut, trends are seen in Virgo (Fig. 5 of Young et al. 2002), Hydra 
A (Figs. 1 \& 3 of McNamara et al. 2000), A133 (Figs. 1 \& 9 of Fujita et al. 
2002), MKW3S (Figs. 1 \& 3 of Mazzotta et al. 2002), 
{\epsscale{1.0}
\plotone{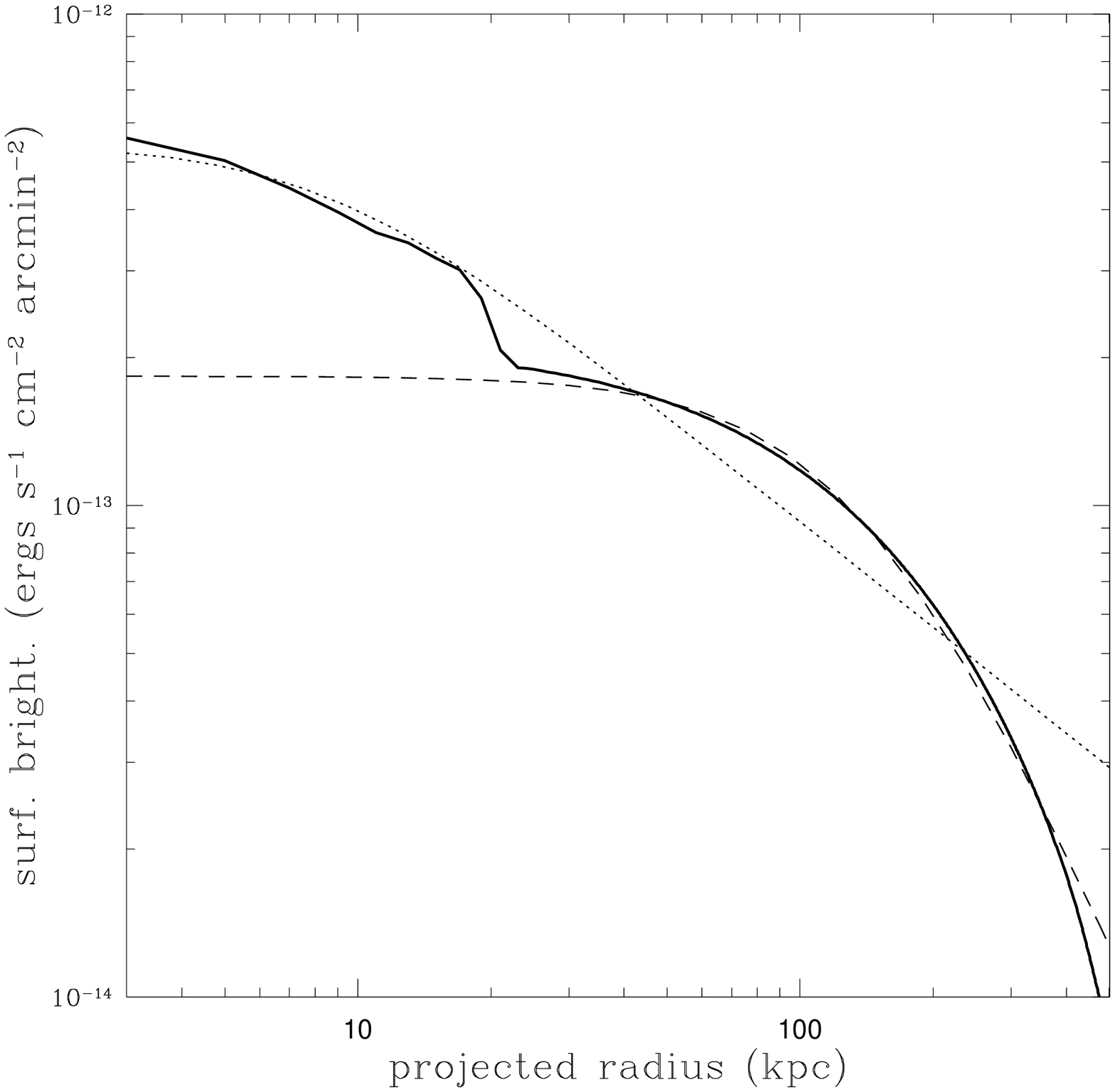}
{Fig. 3. \footnotesize
Predicted bolometric surface brightness profile of the cluster displayed in
Fig.1.  The thick solid line is the resulting profile after the bubbles have been placed
in the cluster.  The dotted line is the best-fit isothermal $\beta$ model, while the
dashed line is the best-fit isothermal $\beta$ model excluding the central 30 kpc.  The
sharp kink at $\approx$ 20 kpc is an artifact of the simplistic geometry we have assumed
for the bubbles. A more realistic geometry would result in a smoother surface brightness
profile.}}
\vskip0.1in
\noindent
Cygnus A (Figs. 1 \& 8 of Smith et al. 2002) and A2199 
(Figs. 2 \& 3 of Johnstone et al. 2002).  Thus, the 
simplistic model we have proposed seems to provide a viable explanation for the gradients 
in these clusters.  However, the model does not appear to be compatible with the {\it 
Chandra} observations of Perseus (Schmidt et al. 2002).  The gradient in that cluster 
extends well beyond the outer radius of the two bubbles situated near the center of the 
cluster.  We note that there are at least two other bubbles at larger radii but they do 
not seem to have bright shells.  Unless the bubbles {\it had} bright shells that somehow 
became dissociated from the cavities and were distributed throughout the ambient ICM, it 
is difficult to see how our model could reproduce the entire temperature gradient of 
Perseus.  Even so, the shells of the two interior bubbles certainly influence the gradient 
near the center of the cluster (note the temperature jump at 50 kpc in Fig. 2 of Schmidt 
et al. 2002).       

What about the surface brightness profiles?  Figure 3 is a plot of the predicted bolometric 
surface brightness profile of the model cluster displayed in Fig. 1.  It is readily 
apparent that the addition of 
bubbles with bright shells results in a sharp peak in the surface brightness profile of 
the model cluster.  This trend holds true for both higher and lower mass model clusters as 
well.  Use of the isothermal $\beta$ model reveals an emission excess at the cluster 
center.  Near the cluster center, the surface brightness has been enhanced by a factor of 
three, which is very similar to what is observed in A2052.  Such excess emission is often 
interpreted as an indicator for the presence of cooling flows (e.g., Blanton et al. 2003) 
but there are no cooling flows in our model clusters.

\section{Discussion}

We have developed a simple toy model that qualitatively reproduces the surface brightness 
and temperature trends of nearby ``cooling flow'' clusters that contain bubbles.  Because 
our models do not have cooling flows, this suggests that the bubbles have significant 
effects on the observed profiles and perhaps explain them entirely (without the need for 
a massive cooling flow).  Without taking into account the cool emission from the bubble 
shells, estimates of the total mass drop out due to radiative cooling would be orders of 
magnitude too high.  Thus, our model potentially explains the longstanding problem of why 
only relatively small amounts of atomic and molecular gas have been found in the centers 
of ``cooling flow'' clusters (e.g., Donahue et al. 2000), at least for some clusters 
(such as A2052).  However, there do exist some ``cooling flow'' clusters that do not have 
bubbles.  Abell 2029, for example, is a seemingly relaxed cluster with a temperature 
gradient that extends out to nearly 260 kpc (Lewis et al. 2002).  This suggests that 
observational features that have come to be characterized as manifestations of cooling 
flows may in fact be due to a wider range of physical phenomena.  As noted earlier, the 
observed properties of Perseus, for example, may be due to several processes, of which 
the bubbles are one.

The results of the present study hinge on the properties of our model bubbles and, in 
particular, their shells.  For the purposes of simplicity, the shell 
properties (i.e., geometry, size, temperature) were {\it chosen} to roughly match the {\it 
Chandra} images of Perseus and A2052, probably the most clearcut cases.  But what physical 
mechanism(s) can give rise to such cool shells?  A number of proposals have recently 
been put forward.  The shells could consist of low entropy gas that was 
lifted by the bubble from the cluster center and cooled through adiabatic expansion 
as the bubble floated to larger cluster radii (e.g., Churazov et al. 2001; Soker et 
al. 2002; Nulsen et al. 2002).  Alternatively, the shells (or shell-like structures) 
could be cool gas from the central cD galaxy that was displaced by a recent merger 
event (Ricker \& Sarazin 2001), the result of instabilities that were induced by the 
interaction between the gas around the cD galaxy and the ICM (Fujita et al. 2002), 
or the result of thermal instabilities that were triggered by radio jets.  Whatever 
the mechanism, the shells should not be regarded as merely re-organized cooling flows, 
since the radiative cooling time of the gas in the shells is apparently larger than 
the age of the bubbles, at least for the limited number of bubbles studied in 
detail to date (Soker et al. 2002; Nulsen et al. 2002).

The cooling time of the gas in the shells may not necessarily be long relative to the age 
of the bubbles for all clusters.  In the absence of a significant source of heating, the 
gas would cool quickly.  This would obviously conflict with the 
lack of X-ray emission lines below $\sim 1$ keV or so in ``cooling flow'' clusters (e.g., 
Peterson et al. 2001).  Thermal conduction has been proposed as a way of explaining the 
lack of 
very cool gas in clusters (e.g., Narayan \& Medvedev 2001; Fabian et al. 2002b), but 
this is over large scales.  In the case of cool shells, conduction would be more 
efficient since it would be acting over smaller scales with a much steeper temperature 
gradient.  In addition, the process of bubble formation itself could help to disentangle 
the magnetic fields in and around the bubbles shells, perhaps allowing conduction to 
proceed near the Spitzer rate.  We suggest the shells could be reheated through 
conduction and eventually disappear when, for example, the jets causing the thermal 
instabilities cease or when the magnetic fields become disentangled enough to allow 
conduction to overwhelm the cooling.  

Ultimately, any detailed model of the ICM must include the natural formation and evolution 
of bubbles with cool shells in realistic galaxy clusters.  High resolution hydrodynamic 
simulations are required and we anticipate that a thorough check of our 
hypothesis will be possible in the not too distant future.  A detailed and explicit 
accounting of the full instrumental response of {\it Chandra}, which has been ignored in 
the present study, should be included in such an analysis.  Hence, we regard the present 
study as a first step towards understanding how bubbles influence the inferred 
properties of the gas in the cores of clusters.  We expect that the results and 
conclusions presented here are generally robust, since the bubble models are based, to a 
large extent, on observations of {\it real} bubbles.  Just how remarkably well this
simplistic model works is, in our opinion, a strong testament to the hypothesis that 
bubbles significantly affect the observed properties of clusters and must 
be taken into account when inferring the physical properties of the ICM.

\vskip0.1in

\noindent We thank the referee, Luca Ciotti, for very helpful comments and suggestions.
I. G. M. is supported by a postgraduate scholarship from NSERC.  A. 
B. is supported by an NSERC operating grant, N. K. is supported by NSF AST-9988146,
NAG5-1203, and NSF AST-0205969 and M. L. B. is 
supported by a PPARC rolling grant for extragalactic astronomy and cosmology at the 
University of Durham.

\end{document}